

\input harvmac.tex
\noblackbox

\def\pmb#1{\setbox0=\hbox{#1}%
\kern-.025em\copy0\kern-\wd0
\kern.05em\copy0\kern-\wd0
\kern-.025em\raise.0433em\box0 }


\Title{}{\vbox{\centerline{Interface Motion in Random Media}
\vskip2pt\centerline{at Finite Temperature}}}


\baselineskip= 16pt plus 2pt minus 1pt\centerline{LEE-WEN CHEN
\footnote{*}
{E-mail:leewen@npac.syr.edu}}
\centerline{and}
\centerline{M. CRISTINA MARCHETTI
\footnote{\dag}
{E-mail: cris@npac.syr.edu}}
\footnote{}
\medskip
\centerline{Physics Department}
\centerline{Syracuse University}
\centerline{Syracuse, NY 13244}
\vskip .1truein

We have studied numerically the dynamics of a driven elastic interface
in a random medium, focusing on the thermal rounding of the depinning
transition and on the behavior in the $T=0$ pinned phase. Thermal
effects are quantitatively more important than expected from simple
dimensional estimates. For sufficient low temperature the creep
velocity at a driving force equal to the $T=0$ depinning force exhibits a
power-law dependence on $T$, in agreement with earlier theoretical
and numerical predictions for CDW's. We have also examined the
dynamics in the $T=0$ pinned phase resulting from slowly increasing
the driving force towards threshold. The distribution of avalanche
sizes $S_\|$ decays as $S_\|^{-1-\kappa}$, with $\kappa = 0.05\pm 0.05$,
in agreement with recent theoretical predictions.

\Date{8/94}

\newsec{Introduction}

The study of the dynamics of driven interfaces in random materials is
relevant to a
wide class of physical problems, from fluid invasion in porous media
\nref\rubio {M.A. Rubio, C.A. Edwards, A. Dougherty and J.P. Gollub,
Phys. Rev.
Lett.  {\bf 63}, 1685 (1989).}\nref\horvath {V.K. Horvath, F. Family and
T. Vicsek, J. Phys. A {\bf 24}, L25 (1991).}\nref\robbins {N. Martys, M.
Cieplak and M.O. Robbins, Phys. Rev. Lett. {\bf 66}, 1058 (1991); N.
Nartys, M.O. Robbins and M. Cieplak, PRB {\bf 44}, 12294
(1992).}\refs{\rubio - \robbins} to the
motion of domain walls in random magnets\nref\grinstein {G. Grinstein and
S.-K. Ma, PRL {\bf 49}, 685 (1982); PRB {\bf 28}, 2588 (1983).}
\nref\huse {D.A. Huse and C.L. Henley, PRL {\bf 54}, 2708 (1985);
D.A. Huse,
C.L. Henley and D.S. Fisher, {\it ibid.} {\bf 55}, 2924 (1985).}\refs
{\grinstein, \huse}.
The interface dynamics in these systems is controlled by the competition
between an external field which exerts a driving force per unit area $F$
on the
interface and the pinning by impurities which impedes the motion.
The interface is often modeled as an elastic medium that is distorted by
disorder, but cannot break.
The dynamics of such elastic interfaces driven through quenched
disorder in the
absence of thermal fluctuations has been studied extensively
both analytically
and numerically\nref\levine{J. Koplik and H. Levine, Phys. Rev. B {\bf 31},
1396(1985); D. A. Kessler, H. Levine and Y. Tu, Phys. Rev. A {\bf 43},
4551(1991).}\nref\reviewt{For a review and list of references, see: J.
Krug and
H. Spohn, {\it Solids Far from Equilibrium : Growth, Morphology and
Defects}, C. Gordriche ed. (Cambridge University press, 1991); G.
Forgacs, R. Lipowsky and Th. M. Nieuwenhuizen, in: Phase Transitions
and Critical Phenomena, vol. {\bf 14}, C. Domb and J. L. Lebowitz,
eds. (Acadamics Press, London(1991)).}\nref\natterman{T. Nattermann,
S. Stepanow, L.-H. Tang and H.
Leschhorn, J. Physique II{\bf 2}, 1483(1992).}\nref\fisher{D.S.
Fisher, PRL {\bf 50}, 1486
(1983); PRB {\bf 31}, 1396 (1985).}\nref\narayan {O. Narayan and D.S.
Fisher, PRB {\bf 48}, 7030 (1993).}
\refs{\rubio - \narayan}.
At zero temperature there is a sharp transition from a sliding
state above a critical driving force $F_T$ to a pinned state below $F_T$. The
transition has been described as a critical phenomenon in terms of
scaling laws
and critical exponents
\refs {\fisher , \narayan}.
The critical exponents depend on the dimensionality of the interface
and of the
embedding space, as well as on the geometry of the quenched disorder. Thermal
fluctuations are expected to round the transition.
Closely related systems that exhibit the same type of nonlinear collective
dynamics and have been studied very extensively are charge density waves
(CDW's) in low dimensional conductors\nref\littlewood {P.B. Littlewood, Phys.
Rev. B {\bf 33}, 6694 (1986); P. Sibani and P.B. Littlewood, Phys. Rev.
Lett. {\bf 64}, 1305 (1990).}\nref\middle {A.A. Middleton and D.S.
Fisher, PRL {\bf 66}, 92 (1991); PRB {\bf 47}, 3530 (1993).}
\nref\myers {C.R. Myers and J.P. Sethna, Phys. Rev. B {\bf 47}, 11171
(1993).}\nref\narayancdw {O. Narayan and D.S. Fisher, PRL {\bf 68}, 3615
(1992); PRB {\bf 46}, 11520 (1992).}[6-11].

The dynamics of weakly pinned flux lines in type-II superconductors
is another
problem in this general class \nref\larkin{A.I. Larkin and Y.N. Ovchinnikov,
J. Low. Temp. Phys. {\bf 34}, 409 (1979).}\refs{\larkin}. A transport
current density ${\bf  J}$ flowing through a type-II superconductor
in the mixed state in
the plane normal to the external field exerts a Lorentz force per unit volume
$F\sim J$ on the vortex lines, which then moves across the current
causing electric
fields proportional to the vortex velocity and hence resistance.
Impurities and
other defects in the material act to pin the vortices and impede their motion.
Point-like material impurities, such as $O_2$ vacancies, yield a quenched
random pinning potential with short-range correlations.

At low temperatures and fields, when the
characteristic pinning energy barriers
for a single vortex line exceed the energy associated with intervortex
interactions, dissipation is controlled by single vortex dynamics. The
problem can then be modeled as that of
a one-dimensional elastic interface (string) driven through a random
medium.
It is important to distinguish between two types of short-range disorder that
are relevant in different physical systems. These are usually referred to as
random-field (or random-force) disorder and random-bond (or random-potential)
disorder. The disorder is of the random-field type for domain walls in
disordered ferromagnets
\nref\skma{G. Grinstein and S.-K. Ma, Phys. Rev. Lett. {\bf 24},
2708(1982).}\refs{\skma}
or fluid interfaces in porous media
\nref\rubio{M. A. Rubio, C. A. Edwards, A. Dougherty and J. P. Collub,
Phys. rev. Lett. {\bf 63}, 1685(1989).}\refs{\rubio}. In this case the
pinning energy is the sum of the contributions from all the impurities in the
area or volume spanned by the invading fluid during its motion. In
contrast, the disorder that controls the dynamics of domain boundaries
in ferromagnets with random exchange interactions
\nref\huse{D. A. Huse and C. L. Henley, Phys. Rev. Lett.
{\bf 24}, 2708(1985)}\refs{\huse}
or flux
lines in type-II superconductors is of the random bond or potential type. In
this case the pinning energy is determined by the instantaneous
position of the
interface and it arises only from the impurities in its vicinity.
The {\it static} behavior, i.e., at finite temperature and zero
driving force, has
been studied extensively and is known to depend strongly on the type of
disorder \refs{\skma,\huse}.
In contrast, Narayan and Fisher recently argued that the critical
behavior of {\it driven} interfaces at zero temperature is essentially
independent of
the type of disorder \refs {\narayan}. Numerical work has not yet addressed
this point conclusively.

In this paper we present  results of numerical simulations of the dynamics
of an elastic string in quenched disorder of the random bond type. Our earlier
studies of this model at zero temperature have focused on the critical
properties at the depinning transition and the  dynamics in the
sliding state \nref\dong{M. Dong, M. C. Marchetti, and A. A.
Middleton, V. Vinokur, Phys. Rev. Lett. {\bf 70},
662(1993).}
\refs{\dong,\natterman}.
We found that at $T=0$ there is a sharp transition at a
threshold force $F_T$ from a state where the string is pinned for
$F<F_T$ to a sliding state for $F>F_T$. The mean velocity vanishes as the
threshold is approached from the sliding state as
$v \sim f^\beta $
with $f=|F-F_T|/ F_T$ and $\beta = 0.24\pm 0.1$
\nref\notea
{ This value is slightly
different from the value $F_T=0.3058\pm 0.0002$ quoted in our earlier
simulations for the same parameters values because we have used a
different form for the pinning potential in the present work. While
the critical exponents describing the transition are not sensitive to
the details of the pinning potential, the numerical value of $F_T$ is
expected to be different for different pinning
parameters}\refs{\notea}.
The data can also be fit
by a form
$v\sim 1 / ln|f|$. The spatial range of velocity correlations in
the sliding state is determined by a correlation length $\xi$ that diverges at
threshold as $\xi \sim f^{ -\nu}$, with $\nu = 1.05 \pm 0.1$. Due to the effect
of random forces the driven interface profile is rough and can be characterized
by a roughness exponent $\zeta$, defined by
\eqn\roughness{<\overline{\Big[u(z,t)-u(0,t)\Big]^2}>\sim |z|^{2\zeta} ,}
where $u(z,t)$ is the instantaneous position of a point on the string.
Our numerical work at $T=0$ yielded $\zeta=0.97\pm 0.05$. Numerical
studies of related lattice models have yielded $\zeta=1.25\pm 0.01$
\nref\leschrzeta{ H. Leschhorn, Phys. Rev. A {\bf 195}, 324(1993).}
\refs{\leschrzeta}.
The continuous elastic model of
interface dynamics was studied recently by Narayan and Fisher \refs
{\narayan} at $T=0$ via a systematic expansion in
$\epsilon = 5-d$, where $d$ is the
dimensionality of the embedding space and $5$ the upper critical
dimensionality. For the case $d=2$ of interest here they found
$\nu = \zeta = 1$
exactly to all order in $\epsilon$ and $\beta = 1-2\epsilon/9 +
{\cal O}(\epsilon{^2})= 1/3+{\cal O}(\epsilon{^2})$. These results
are consistent with those obtained from our numerical work.

We have now extended our work in two ways. First we have
considered the string dynamics in
the presence of both quenched disorder and thermal fluctuations.
Secondly, we have investigated the properties of the zero temperature
pinned phase below threshold.
As expected, thermal fluctuations round the depinning transition and
replace it by a smooth crossover. For temperatures well below a
characteristic depinning temperature $T_{dp}\sim (\Delta K
R_p^3)^{1/3}$, where $\Delta$ is the strength of the quenched disorder
(of range $R_p$) and $K$ the string elastic constant, the
temperature dependence of the mean velocity at the zero
temperature threshold $F_T$ scales with temperature as $v(F_T,T)\sim
T^{\beta/\tau}$, where $\beta$ is the $T=0$ depinning exponent and
$\tau$ is a nonuniversal exponent, with $\tau={3\over 2}$ for
continuous pinning potentials (see Figure 1). This scaling form was
predicted some time ago \refs{\narayancdw} by D. S. Fisher
through a mean field
theory for the Fukuyama-Lee-Rice model of CDW's. It was also recently
verified numerically for both continuous and discrete CDW models in two
and three dimensions \nref\middlet{A. A. Middleton, PRB \underbar{45},
9465(1992).} \refs{\middlet}. It is not {\it a priori}\ obvious
that interfaces with
short-range pinning potentials should behave like models of CDW's,
where the pinning potential is periodic - and therefore correlated -
along the direction of motion. In fact the critical exponents
describing the $T=0$ depinning transition are different in these two
classes of models. On the other hand, our results indicate that
in both cases the dynamics at
finite temperatures occurs via thermally induced hopping over small
energy barriers and is insensitive to the details of the pinning
potential, provided the time scale of the thermal hop is sufficiently
short. The mechanism for destabilization of the soft modes of the
system is the same for models with short-range pinning potentials and
CDW's and the mean field argument proposed by Fisher applies in both
cases. The details of the pinning potential only enter
through the $T=0$ depinning exponent, $\beta$.
The string dynamics for $T>T_{dp}$ is qualitatively different and the
scaling just described fails. This difference is clarified by
the high force perturbation theory discussed in Section 3.
Numerical studies of interface dynamics in $1+1$ dimensions at finite
temperature
were
carried out recently by Kaper et al.\nref\kaper{H. Kaper, G. Leaf,
D. Levine, V. Vinkur, Phys. Rev. Lett. \underbar{71},
3713(1993)} \refs{\kaper}. Our work
on thermal effects complements this earlier studies by
focusing on the scaling of velocity with temperature for $T<T_{dp}$,
that was not discussed in \refs{\kaper}, and on the qualitative
difference in the string dynamics at low $(T<T_{dp})$ and high
$(T>T_{dp})$ temperatures.

We have also investigated the properties of the zero temperature
pinned phase below threshold. The response of the string when
the driving force is increased towards threshold from below
is dominated by
localized forward jumps or ``avalanches''. The number distribution of
avalanche sizes $S_\|$ in the string direction, $D(S_\|;f)$, was
conjectured to obey a scaling form near threshold
\refs{\narayan}, characterized by a diverging correlation
length $\xi\_\sim (F_T-F)^{-\nu\_}$,
\eqn\dsparr{
D(S_\|;f)={1\over S_\|^{1+\kappa}}\hat D(S_\|/\xi_{-}),
}
where $D(S_\|;f)dS_\|df$ is the number of avalanches of diameter
between $S_\|$ and $S_\|+dS_\|$ that occur when the reduced
force is changed from $f$ to $f+df$. Also, $\hat D(x)$ is a scaling
function that decays rapidly for $x>>1$.
Using an $\epsilon$ expansion in $5-\epsilon$ dimensions, Narayan and
Fisher predicted $\kappa=0$ and $\nu\_=\nu=1$, with $\nu$ the
correlation length exponent when the transition is approached from
above threshold \refs{\narayan}.

We have studied the scaling of
avalanches numerically by evaluating the distribution ${\cal D}(S_\|)$
of avalanche
sizes $S_\|$ obtained in response to a small increase of the
driving force for all forces below $F_T$,
\eqn\intspar{
{\cal D}(S_\|)=\int_{-1}^0dfD(S_\|;f)\sim {1\over S_\|^{1+\kappa '}},}
and we find
$\kappa'=1.0\pm 0.2$. The exponents $\kappa$ and $\kappa '$ are
related by
\eqn\exprel{
\kappa '=\kappa +{1\over\nu\_}.
}
We find $\kappa '=1.0\pm
0.2$ and $\kappa = 0.05\pm 0.05$, corresponding to $\nu\_=1.13\pm
0.30$, consistent with the theoretical values.
The value $\kappa=0$
indicates that avalanches of large size are likely and
the description of the interface or as an elastic string breaks down.
This is also consistent with our result that the roughening exponent
$\zeta$ defined as $A\sim S_\|^{1+\zeta}$, where $A$ is the area of
an avalanche, is
$\zeta=1.0\pm 0.05$. As discussed by Coppersmith
\nref\smithcdw{S. N. Coppersmith, Phys. Rev. Lett. {\bf 65},
1044(1990).}
\refs{\smithcdw}
in the context of
CDW, large strain may develop in the $d=2$ case of interest here. In a
real system the string will relax these strains by pinching off a
vortex loop around an impurity, a mechanism that is excluded in our
model. Overhangs and loops could modify qualitatively both the
dynamics in the pinned region and the $T=0$ depinning transition.
Further studied are needed to address their role.

The remainder of the paper is organized as follows. After briefly
describing the model in section 2, we discuss simple dimensional
estimates and the results of the high field perturbation theory in
section 3. The numerical result at finite temperature are presented in
section 4, while the scaling of avalanches in the $T=0$ pinned state
is discussed in section 5.

\newsec{The Model}

The specific model we have studied is an elastic string
embedded in two dimensions and on the average aligned with the $z$
direction (in superconductors this is the direction of the applied magnetic
field). We describe the interface by its displacement $u(z,t)$ in
the direction
of the applied force per unit length, $F$. The displacement is assumed
to be a
single-valued function of $z$ at any time $t$, i.e., we ignore overhangs
in the
string and vortex loops that could be pinched off from it during the motion.
In the absence of driving force the Hamiltonian of the string is the sum of
the elastic energy of the string and a pinning potential from random
impurities
in the medium,
\eqn\energy{H=\int^L_0\Big\{{K\over 2}({\partial u\over\partial
z})^2+ V(u(z,t);z)\Big\},}
where $L$ is the size of the system in the $z$ direction and
$K$ the elastic constant.
Assuming purely relaxational dynamics, the equation of motion
for the string in
the presence of a constant driving force $F$ per unit length in the
transverse direction is given by
\eqn\motion{\eqalign{\gamma\partial_tu(z,t)=&-{\delta H\over\delta
u}+F+\eta(z,t)\cr
=&K\partial_z^2u(z,t)+F_p(u;z)+F+\eta(z,t),}}
where $\gamma$ is a friction coefficient, $F_p(u;z)=-\partial
V(u;z)/\partial u$ is the pinning force per unit length,
and $\eta (z,t)$ is a Gaussian
correlated Langevin force per unit length describing thermal noise, with
$<\eta (z,t)>=0$ and correlations
\eqn\noise{<\eta(z,t)\eta(z',t')>=2\gamma T\delta(z-z')\delta(t-t').}
The angular brackets $< \cdot\cdot\cdot>$ denote the thermal average.
In the numerical model the pinning potential is written
explicitly in terms of the interaction of the string with $N_p$
short-ranged pinning centers randomly distributed at positions
$\vec {R}_i$ in the plane,
\eqn\pinning{V(u;z)=\sum_{i=1}^{N_p}U(|\vec {r}-{\vec {R}}_i|),}
where $\vec {r} = (u(z,t),z)$. The interaction
$U(|\vec {r} - \vec {R}_i|)$ of the string with the $i^{th}$ pin is
approximated by a potential well centered at the pin location
$\vec{R}_i$ of finite range $R_p$ and maximum depth $U_0$. The
pins are uniformly distributed with areal density $n_p$. When
the mean pin spacing $1/ \sqrt{n_p}$ is large compared to the
potential range $R_p$, the pinning potential can also be described as
a continuum gaussian random variable, with mean and correlations given
by,
%
\eqn\quenchpot{\overline{V(u;z)V(u';z')}\sim\Delta\delta(u-u')\delta(z-z'),}
where
\eqn\disorder{\Delta\approx (U_0R_p)^2n_pR_p^2\{1+{\it O} (n_pR_p^2)\}}
and the overbar denotes an average over disorder realizations.

The overall motion of the string is described by a ``center of
mass'' velocity, defined as
\eqn\cmvel{v_{cm}(F,t)={1\over L}\int_0^Ldzv(z,t),}
where $v(z,t) = \partial_{t} u(z,t)$ is the instantaneous velocity
of a point on
the string. The center of mass velocity is a fluctuating quantity since it
depends of both the random positions of pins and the thermal noise.
The average or drift velocity of the string is given by
\eqn\driftvel{v(F)=\overline{<v_{cm}(F,t)>}.}
In the numerical calculation the average over different realizations of
disorder (denoted by the overbar) is performed by averaging over time,
since as
time evolves the string samples different impurity configurations.

\newsec{Dimensional estimates and perturbation theory}
Considerable insight on the behavior of the driven string can be
gained by simple dimensional estimates and by a perturbation theory in
the pinning force. Most of the results discussed in this section have
been obtained elsewhere for a general d-dimensional interface
\nref\longrev{Vortices in High Temperature Superconductors, G.
Blatter, M. V. Feigel'man, V. B. Geshkenbein, A. I. Larkin, V. M.
Vinokur.}\refs{\longrev}. It is, however, instructive to summarize
them here for the $1+1$ dimensional case of interest.

The relative importance of thermal fluctuations and quenched disorder in
governing the {\it equilibrium} properties of elastic interfaces,
in the absence of external drive, has been studied extensively.
Thermal fluctuations always dominate
at small length scales. In this regime they are responsible for small amplitude
vibrations of the interface within a given pinning well and therefore lead
to a smoothing of the pinning potential. At large length scales the importance
of thermal fluctuations depends on dimensionality. For a one dimensional
interface in $d=2$ the disorder is always dominant at large length scales.
This is because the value of the roughness exponent $\zeta_{eq}$ of an elastic
string disordered only by random bond quenched disorder, $\zeta_{eq} (d=2) =
2/3$, is larger than the corresponding value for a thermal string in
equilibrium, $\zeta_{th}=1/2$.

A disordered elastic interface in the absence of thermal
fluctuations is characterized by the Larkin-Ovchinnikov (LO)
collective pinning length,
$L{_c}$. This is estimated by dimensional analysis by considering
the energy fluctuation associated with displacing
a segment of string of length $L$ by a transverse
distance $R_p$ across a single pinning well,
\eqn\freefluct{\delta F =K {R_p^2\over
L}+(U_0R_p)\sqrt{n_pR_pL}.}
The $LO$ pinning length is obtained by minimizing \freefluct , with
the result \  $L_c\approx R_p\big (K/U_0\sqrt{n_pR_p^2}\big )
^{1/3}$.
When $L_c >> R_p$, the string is pinned collectively by
many pins. It is this weak pinning regime that is relevant for flux lines in
superconductors. When thermal fluctuations are important the
mean thermal displacement of the string
exceeds $R_p$ and the string experiences a
random potential averaged over the length of its root mean square
thermal excursion about equilibrium, $<u_{th}^2>^{1/2}$. The temperature
dependent pinning length is estimated by
considering the energy associated with a fluctuation of length $L$ and
transverse distance $<u_{th}^2>^{1/2}$,
\eqn\freethermal{
\delta F = K{<u_{th}^2>\over L}+U_0R_p\Big[n_pR_p
L{R_p\over<u^2_{th}>^{1/2}}\Big]^{1/2} ,}
where $<u_{th}^2>$ is determined by requiring $K
<u^2_{th}>/L\sim T$.
We find $L_c(T)\approx L_c(T/T_{dp})^5$,
where the depinning temperature $T_{dp}$ is defined by
$<u_{th}^2>\approx L_c(T)T/K=R_p^2$
, which yields $T_{dp}\approx (U_0R_p)^{2/3}(n_pR_p^3K)^{1/3}$.
The temperature-dependent pinning length is defined by interpolating
between the $T=0$ LO length and the finite temperature
result,
\eqn\lcot{
L_c(T)=L_c[1+(T/T_{dp})^5].
}
A more rigorous derivation of these results can be found in
\refs{\longrev}.

Considerable insight on the competing roles of quenched disorder and thermal
noise can be gained by considering the string dynamics at
large velocities, where the disorder can be treated as weak. For
$F>>F_T$, the effect of pinning is negligible and the string
advances uniformly, with $v\approx v_0 = F/ \gamma$. The deviations from this
asymptotic behavior can be studied by a perturbation theory in
$F/F_p$ that was
introduced first by Schmid and Hauger and by Larkin and Ovchinnikov
for $T=0$ and recently extended by
Vinokur et al.\nref\vinokper{V. M. Vinokur, V. B. Geshkenbein, A. I.
Larkin, V. M. Feigel'man, Sov. Phys. JETP \underbar{73}(3),
610(1991).}
\nref\feigelman{M. V. Feigel'man, Sov. Phys. JETP {\bf 58}, 1076(1983).}
\refs{\vinokper,\feigelman}
to incorporate thermal fluctuations. To carry out the perturbation
theory it is convenient to consider a frame of reference moving at the
mean velocity $v$ of the string. The instantaneous position of a point
on the string is then written as $u(z,t)=vt+u_p(z,t)+u_{th}(z,t)$,
where $u_p(z,t)$ is the deformation due to quenched disorder, treated
as small, and $u_{th}(z,t)$ is the contribution from thermal
fluctuations to the displacement from the uniform sliding state. It is
defined as the solution of,
\eqn\thermalmot{
\partial_t u_{th}(z,t)=K\partial_z^2 u_{th}(z,t)+\
\eta(z,t).
}
The mean velocity is then given by
\eqn\avgfp{
\delta v={1\over\gamma}<\overline{F_p(vt+u_p+u_{th};z)}>,
}
with $\delta v=v-v_0$ and
\eqn\pinmot{
\gamma\partial_t u_p(z,t)=K\partial_z^2
u_p(z,t)+F_p\big(vt+u_p(z,t)+u_{th}(z,t)\big). }
The right hand side of Eq. \avgfp\ is evaluated in perturbation theory
by treating $u_p$ as small.
The first order correction
to the average string velocity as compared to the asymptotic
value $v=v_0$ is
given by the self-consistent solution of the equation
\eqn\velpert{{\delta v\over v}={\Delta\over \gamma}
\int_{-\infty}^{+\infty}{dk\over 2\pi}k^3|p(k)|^2
\int_0^\infty dt{\sin kvt\over v}G(0,t)S_s(k,t),}
where $p(k)$ is the single-vortex
form factor that provides a
short-length scale cutoff at $k\sim 1/R_p$,
with $k$ the  wavevector in the direction
of motion.
Also, $G(0,t)$ is the vortex Green's function evaluated at $z=0$,
given by
\eqn\green{G(0,t)=\Theta(t)\int^{'}{dq\over
2\pi}{1\over\gamma}e^{-q^2K t/\gamma}.}
The prime on the integration over the wavevector $-\infty <q<+\infty$
in the $z$ direction
denotes a short distance cutoff at $|q| \sim
2\pi/R_p$ which controls the crossover to single particle behavior at
short times.
Thermal fluctuations are responsible for the appearance in Eq.
\velpert\
dynamical structure factor $S{_s}(k,t)$ of a single vortex in the
absence of disorder, given by
\eqn\structure{S_s(k,t)=< e^{ik[u_{th}(z,t)-u_{th}(z,0)]} >
\approx e^{-k^2<[u_{th}(z,t)-u_{th}(z,0)]^2 >/2}.}
In the absence of thermal fluctuations $S{_s}(k,t)=1$ and the cutoff $k_0$ on
the $k$-integration in Eq. \velpert\ is provided by the form
factor $p(k)$, i.e.
$k_0\sim1/R_p$. At finite temperature this cutoff is replaced by
$k_0\sim [R_p^2+<u_{th}^2(t)>/2]^{-1/2}$.

By using the fluctuation-dissipation theorem the mean square thermal
displacement can be expressed in terms of the Green's function given in Eq.
\green , with the result
\eqn\msquare{<u_{th}^2(t)>={2T\over K}\int^{'}{dq\over 2\pi}{1\over
q^2}\Big(1-e^{-K q^2t/\gamma}\Big).}
The $q$-integration on the
right hand side of Eq. \msquare\ is easily carried out, with the result
\eqn\msthermal{\eqalign{<u_{th}^2(t)>&\approx
{T\over\pi\sqrt{K\gamma}}\sqrt{t},\qquad t>>t_0,\cr
&\approx {4T\over \gamma R_p}t,\qquad t<<t_0,}}
where $t_0 = R_p^{2}/D_0$, with $D_0=K/\gamma$ the diffusion
constant of a point vortex, is the time scale for diffusion across
the range $R_p$ of the pinning potential. For $t<<t_0$ different bits
of string of length $\leq R_p$
are essentially uncorrelated along the $z$ direction and one
recovers single-particle behavior, $<u_{th}^2(t)>\sim t$.

As discussed earlier, at large times or long length scales disorder is
always important and the long time divergence of Eq. \msthermal\
is cut off by disorder. The effect of disorder can be approximately
incorporated in Eq. \msquare\ by introducing a large distance cutoff at
$q\sim 2\pi/L_c(T)$ in the q-integration.
Neglecting for simplicity the crossover to single particle behavior, we
obtain three distinct regimes
\eqn\msfull{
\eqalign{
<u_{th}^2(t)>\;& <
R_p^2,\qquad t_0<t<t_{ph}=t_0\Big(R_pK/T\Big)^2,\cr
<u_{th}^2(t)>\;&\approx {T\over\pi}\sqrt{{t\over K\gamma}},\qquad
t_{ph}<t<t_d=t_{ph}\Big(T/T_{dp}\Big)^{12},\cr
<u_{th}^2(t)>\;&\approx {TL_c(T)\over K},
\qquad t>t_d.}}
We have introduced here two new time scales. The time $t_{ph}$
characterizes the
time scale for small phonon-like vibrations of the string within the
potential well of a single pin, while $t_d$ is the time scale where disorder
becomes dominant. The intermediate regime in Eq. \msfull\ only
occurs provided
$t_{ph} < t_d$ which corresponds to $T > T_{dp}$.

We now proceed to approximately evaluate the correction
$\delta v$ given in Eq.
\velpert\ . The right hand side of Eq. \velpert\ can be reduced to a
one-dimensional integral that can be evaluated numerically. It is, however,
more instructive to simply carry out a dimensional estimate of the integral.
Neglecting for now the short distance cutoff in the $q$-equation in
Eq. \green\ ,
we obtain $G(0,t)=\Theta(t)\sqrt{\gamma/\pi K t}$.
Inserting this on the right hand side of Eq. \velpert\ and
noting that the main contribution to the time integration comes from $kvt \sim
1$, we obtain
\eqn\deltav{
{\delta v\over v}={\Delta \over
\sqrt{\pi\gamma K }}
\int_{-\infty}^{+\infty}{dk\over 2\pi}k^4
\int_0^{1/kv} dt\sqrt{t}e^{-k^2[R_p^2+<u_{th}^2(t)>/2]}.}
As the driving force $F$ - and therefore the string mean velocity $v$
- increases, the time cutoff $1/kv$ decreases. Inserting on the right
hand side of eq. \deltav\ the form for $<u^2_{th}(t)>$ from Eq.
\msfull\ appropriate to each time regime and carrying out the
integration, we obtain
\eqn\asympt{
\eqalign{
 {\delta v\over v} &\sim
{\Delta\over\gamma R_p^3\sqrt{K\gamma R_p}}{1\over v^{3/2}}
, \qquad\qquad t_0<{R_p\over v}<t_{ph},\cr
 {\delta v\over v} &\sim
{2\Delta\over 15}\Biggl({2^7K^2\gamma^2\over\pi T^7}\Biggl)^{1/3}
{1\over v^{1/3}}
, \qquad t_{ph}<{R_p\over v}<t_d,\cr
 {\delta v\over v} &\sim
{2\Delta\over 15}{1\over\sqrt{\pi K\gamma}}\Biggl({\epsilon\over
TL_c(T)}\Biggl)^{7/2}
{1\over v^{3/2}}
, \qquad t_d<{R_p\over v},}}
or
\eqn\fcrossover{
\eqalign{
1-{v\over v_0}&\sim F^{-3/2},
\qquad {T^2\over K R_p^3}<F<{K\over R_p},\cr
1-{v\over v_0}&\sim F^{-1/3},
\qquad {T_{dp}^2\over K R_p^3}\Big({T_{dp}\over T}\Big)^{10}
<F<{T^2\over K R_p^3},\cr
1-{v\over v_0}&\sim F^{-3/2},
\qquad F<{T_{dp}^2\over K R_p^3}\Big({T_{dp}\over T}\Big)^{10}.
}
}
For $F>K/R_p$, corresponding to the case where the time cutoff is
realized at times shorter than $t_0$, the string is
sliding so fast that correlations among different string elements have
no time to develop and one recovers single particle behavior, with
$\delta v/F\sim F^{-2}$.
The results summarized in Eqs. \fcrossover\ show that the shape of the
$v$-$F$ curve is qualitatively different at high and low temperatures.
The perturbation theory yields $v/F\sim 1-
CF^{-\alpha}$, where C is a constant that depend on temperature and the
value of $\alpha$ in the various regimes can be inferred from Eqs.
\fcrossover . The curvature of the $v$-$F$ curve is determined by
${d^2v\over d^2F}\sim -\alpha (\alpha -1)F^{-\alpha -1}$ and is
positive if $\alpha < 1$ and negative if $\alpha > 1$. If $T<<T_{dp}$
the intermediate region described by the second of Eqs. \fcrossover\
does not occur and the
perturbation theory yields $\alpha=3/2$ in the entire region
where perturbation theory applies, up to $F\sim K/R_p$.
The $v$-$F$ curve has a negative
curvature in this regime and thermal effects only affect the
coefficient of the correction $\delta v$.
When $T>T_{dp}$ there is an intermediate region
described by the second of Eqs. \fcrossover\ where $\alpha=1/3$ and
the $v$-$F$ curve
has a positive curvature. This behavior is apparent in our data
described in the following section(see Fig. 2).
\newsec{Numerical Results at Finite Temperature}

We have integrated numerically the discretized version of the equation
of motion \motion\ for a string composed of discrete elements, each of
dimensionless size $R_p$ in the $z$ direction. Here $R_p$ is chosen as
the unit of lengths. All forces are measured in units of the string
tension $K$. The string elements interact via nearest neighbor
elastic forces and are constrained to move only in the direction of
the driving force. This model is appropriate for flux lines in layered
superconductors, where each flux line can be thought of as a stack of
interacting two-dimensional ``pancake'' vortices residing in the
$CuO_2$ planes. Periodic boundary conditions are imposed in the
$z$ direction.
Each string element is subject to attractive potential wells of finite
size $R_p=1$ and maximum depth $U_0$ in both the $z$ and $u$ directions,
centered at the randomly
distributed pin locations.
The displacements  of the discrete string elements
are treated as continuous variables and the coupled
equations of motion are integrated by a fourth order Runge-Kutta
algorithm with a time step much smaller than the typical time to cross
a single potential well (typically $\Delta t\sim 0.1$, where time is
measured in units of $t_0=R_p^2/D_0$). As in our $T=0$ simulations, we
have been able to obtain reliable data at very small velocities thanks
to our ``pinning cells'' method. The method consists in dividing the
plane in ``pinning cells'' of dimension $R_p$. At $T=0$ the string can
only move forward and each section of the string only ``sees'' the
disorder in the pinning cell that neighbors it in the forward
direction of motion. Thermal fluctuation can also kick the string to
move backward, in the direction opposite to that of the driving force.
At every iteration each string element needs to know the disorder in
the cells that it has left behind, since these may be revisited. We
have developed an algorithm to store at each step an adjustable number of
``past'' pinning cells for each section of the string, in addition to
the ``future'' cell. The list of past cells is updated in parallel at
each time step. For driving forces within $10\%$ of the zero temperature
threshold, the length of the simulation usually exceeds $10^6$ time
steps, while shorter simulation times gave good averages for forces
further from threshold. We have investigated systems of size from 256
to 16384. The computations were performed on the Connection Machine
CM-2 and CM-5. In our dimensionless units the parameters of the model
are the dimensionless pinning force $\tilde F_p=U_0/K$, the
dimensionless areal density of pins, $\rho=n_pR_p^2$, and the
dimensionless driving force $\tilde F=FR_p/K$. In these units
the temperature is measured in units of $KR_p$ and the depinning
temperature is given by $\tilde T_{dp}=T_{dp}/K =\tilde
F_P^{2/3}\rho^{1/3}$ and the LO collective pinning length is
$\tilde L_c=\tilde F_p^{-1/3}\rho^{-1/6}$, yielding a dimensionless
estimate for threshold force $\tilde F_T=\rho\tilde F_p$ and $\tilde
F_T=\rho^{2/3}\tilde F_P^{4/3}$ for strong and weak pinning,
respectively. We will always refer to the dimensionless quantities
below and drop the tilde to simplify the notation.

Our simulations were carried out for $F_p=1$ and $\rho=0.1$, yielding
$L_c=1.5$ and $T_{dp}=0.46$. The dimensional estimate for the
threshold force gives $F_T=0.1$, which is consistent with our result
$F_T=0.2435\pm 0.005$ obtained from the simulations.
The mean velocity of the string is shown in Fig. 2 for various
temperatures and system sizes. For very large driving forces the
effect of pinning is negligible and the string advances uniformly,
with $v\approx F$. For $T\neq 0$ there is no sharp
transition and the velocity at low driving forces is small but finite.
The velocity versus driving force ($v$-$F$) curve always exhibits
a tail with positive curvature below the
$T=0$ threshold. At higher driving forces the curvature of the $v$-$F$
curve changes sign and eventually approaches the asymptotic limit
$v\sim F$. At $T=1.0\times 10^{-2}$, well below our dimensional
estimate of $T_{dp}\approx 0.46$, thermal effects have already washed out
completely the depinning transition. Thermal effects are
quantitatively more pronounced than expected from dimensional
estimates.

We distinguish three regimes characterizing the string
response at finite temperature. At high driving forces the deviation
from the asymptotic behavior $v\sim F$ are well described by the
perturbation theory discussed in section 3. If $T<<T_{dp}$ our
simulation agree with $\delta v/F\sim F^{-3/2}$(with a crossover to
$\delta v/F\sim F^{-2}$ at very large driving force) and
the $v$-$F$ curve has a negative curvature
in this regime. At higher temperature the
$v$-$F$ curve shows an intermediate region with positive
curvature where $\delta v/F\sim F^{-1/3}$.
The qualitative difference in the shape of the $v$-$F$ curve
at high and low temperature is consistent with the results of the
perturbation theory discussed at the end of Section 3. The value of
$T$ above which thermal effects change qualitatively the response is,
however, much lower than the value obtained for $T_{dp}$ from
dimensional estimates.
Finally, at very high forces, $1-v/F\sim F^{-2}$.
This crossover is controlled by the
range of the pinning potential correlations in the $z$ direction.

For driving forces within a few percent of the zero temperature
threshold and sufficiently low temperature,
the mean velocity exhibits the scaling behavior proposed by
Fisher in the context of CDW's. He argued that thermally-induced
``hops'' over pinning energy barriers are analogue to jumps resulting
from ramping up the force at $T=0$. The velocity evaluated at the
$T=0$ threshold is found to scale with temperature as $v(F=F_T,T)\sim
T^{\beta/\tau}$, with $\beta=0.24$ and $\tau=3/2$,
as shown in Fig. 1. In addition in the region of low
$T$ and $F$ close to $F_T$, our data can be fit to the scaling form
proposed by Fisher
\eqn\scalt{
v(f,T)\sim T^{\beta/\tau}B(fT^{-1/\tau}),
}
as shown in Fig. 3. Here $B(x)$ is a scaling
function that behaves as $B(x)\sim x^\beta$ for $x\rightarrow\infty$.
This scaling fails, however, at higher temperatures.

Finally, at very small driving forces ($F<<F_T$) and low temperatures
($T<<T_{dp}$) the dynamics occurs via creep over pinning
energy barriers and $v\sim e^{-U(F)/T}$. The pinning energy barrier
$U(F)$ has been predicted to diverge when $F\rightarrow 0$ as
$U(F)\sim F^{-\mu}$. The creep exponent $\mu$ has been
estimated by dimensional analysis as $\mu=1/4$. It should, however,
be noticed that this result was
obtained by assuming that the roughening exponent of the string has
the equilibrium value $\zeta_{eq}=2/3$.
Our numerical results in this region are not inconsistent with a small
value of $\mu$, as shown in Fig. 4, but are insufficient
to either confirm or discard the assumption that the creep dynamics
is controlled by the single diverging energy scale $U(F)$ and
to determine
the creep exponent conclusively. A different approach may be needed
to address this point.
\newsec{Avalanches in the $T=0$ pinned state}

In this section we discuss the critical behavior of
the interface at zero temperature in the pinned region, as the
threshold $F_T$ is approached from below by slowly increasing the
driving force. As discussed recently by Narayan and Fisher
\refs{\narayan} for the interface problem and in more
detail by Narayan and Middleton
\nref\middletava{O. Narayan, A. A. Middleton, Phys. Rev. B
\underbar{49}, 244(1994)}\refs{\middletava} for CDW's, local
instabilities occurs as $F$ is increased towards threshold resulting
in ``avalanches'' of various size.
The notion of avalanches was introduced to
describe the large response to a local perturbation in models
exhibiting self-organized criticality (SOC)\nref\sneppen{K. Sneppen,
Phys. Rev. Lett. {\bf 69}, 3539(1992).}\refs{\sneppen}.
Here we
discuss avalanches obtained in response to a small
increase of the driving force, which provides a global perturbation that
affects equally all the discrete string elements. This perturbation, if
sufficiently small, will, however, only trigger instabilities locally.
The resulting response consists of discontinuous local jumps of
portions of the interface (i.e., avalanches), with a distribution of sizes.
The size of an avalanche can be defined in terms of its area (or
moment)$A$ or of its diameter $S_\|$ in the direction of interface.
We start the system in a pinned
configuration at a driving force $F<F_T$. Increasing the reduced force
$f=F/F_T-1$ from f to $f+df$ triggers a forward jump of sections of
the string, until a new metastable pinned state is reached. The
avalanche is characterized by the total area $A$ swept by the string
as a result of the increase in driving force,
\eqn\avaa{
A=\int_0^{S_\|}dzu(z).
}
Assuming $u(z)\sim z^\zeta$,
with $\zeta$ a roughening exponent, the area and diameter of the
avalanche are related by
\eqn\areascaling{
A\sim S_\|^{1+\zeta}.
}
Figure 5 shows a plot of the area of the avalanches versus their
diameter. The straight line has slope 2 and we find $\zeta=1\pm 0.05$.

Fisher and Narayan have conjectured that the distribution of
avalanche sizes near threshold obeys the scaling form
given in Eq. \dsparr \refs{\narayan}.
Using Eq.\areascaling , one can also immediately obtain
the scaling form for the number distribution of
avalanche areas, $D_A(A;f)$, given by,
\eqn\dascalf{
D_A(A;f)={1\over A^{1+\kappa_A}}\hat D(A^{1/(1+\zeta)}/\xi\_),}
with
\eqn\kappaA{
\kappa_A={\kappa\over(1+\zeta)}.}
Rather than considering the distribution of avalanches at a fixed
force $f$,
we have evaluated numerically the distribution of avalanche
areas and diameters integrating over all
driving force below $F_T$. The total distribution of avalanche size
is given in Eq. \intspar . The corresponding distribution of avalanche
areas is given by
\eqn\intarea{
{\cal D}_A(A)=\int_{-1}^0dfD_A(A;f)\sim {1\over A^{1+\kappa_A'}}.
}
The corresponding exponents are related to the exponents defined in
\dsparr and \dascalf by
\eqn\primedexp{\eqalign{
\kappa ' &=\kappa+{1\over\nu\_}\qquad ,\cr
\kappa_A' &=\kappa_A+{1\over \nu\_(1+\zeta)}\qquad .}}
To generate avalanches, we start with the string in a metastable
pinned configuration well below $F_T$ and study the response to a
small increase $\Delta F$ of the force (we have used $\Delta F=1\times
10^{-4}$ in most of our simulations).
The increase in force triggers local avalanches which are recorded,
until the string reaches a new pinned configurations. The procedure is
then repeated by stepping up again the
force of an amount $\Delta F$ until $F_T$ is reached.
If, however, the string
slides as a whole in response to the small perturbation, we
discard this event, return to the original pinned configuration and
repeat the procedure with a smaller $\Delta F$. These ``giant
avalanches'' are discarded because they are a finite-size effect,
characteristic of the response of finite systems above threshold.
The distributions are shown in Figs. 6a and 6b for two system sizes.
They exhibit a power-law decay with $\kappa '=1.0\pm 0.2$ and
$\kappa_A'=0.5\pm 0.15$.

In order to gain further insight in the ``shape'' of the avalanches,
we define the number distribution $D_u(\Delta u;f)$ of the
local displacement advances $\Delta u$ following the ramping of the
force. More precisely $D_u(\Delta u;f)d(\Delta u)df$ is the number of
displacement advances between $\Delta u$ and $\Delta u+d(\Delta u)$
that occur when the reduced force is increased from $f$ to $f+df$.
To develop a scaling ansatz for $D_u(\Delta u;z)$ we proceed as follows.
Let $n(\Delta u|A)d(\Delta u)$ be the conditional number of
displacements between $\Delta u$ and $\Delta u+d(\Delta u)$ that
occurs within a given avalanche of fixed area $A$. We assume that the
transverse shape of the avalanche (in the direction of motion) can be
characterized by a single length scale $S_\perp$, defined as
\eqn\sperp{
S_\perp=A/S_\|.
}
then conjecture a scaling ansatz for $n(\Delta u|A)$ of the form
\eqn\condprobs{
n(\Delta u|A)={S_\|\over S_\perp}\hat n(\Delta u/S_\|),
}
where $\hat n(x)$ is a scaling function that depends on the detailed
shape of the string and the prefactor in Eq. \condprobs\ is
determined by the normalization condition
\eqn\normalp{
\int_0^{S_\perp}d(\Delta u)n(\Delta u|A)=S_\|.
}
The distribution of displacements $D_u(\Delta u; f)$ can then be written
as
\eqn\udistub{
D_u(\Delta u;f)=\int dA D_A(A;f) n(\Delta u|A).
}
By inserting Eq. \condprobs in Eq. \udistub and making use of Eq.
\areascaling and \sperp , we obtain
\eqn\gdu{
D_u(\Delta u;f)\sim {1\over \Delta u^{1+\kappa_u}},
}
with
\eqn\kappau{
\kappa_u = -{1\over\zeta}+\kappa_A\big(1+{1\over\zeta}\big).
}
Finally, the number distribution of the local displacements $\Delta u$
integrated over all forces up to $F_T$ decays as
\eqn\intdelu{
{\cal D}_u(\Delta u)\sim {1\over \Delta u^{1+\kappa_u'}},}
with
\eqn\kappaup{
\kappa_u'=\kappa_u+{1\over\zeta\nu\_}.}
The distribution ${\cal D}_u(\Delta u)$ can be evaluated with excellent
statistics and is shown in Fig. 7.
We find $\kappa_u'=0.0\pm 0.10$.

The primed exponents $\kappa ',\kappa_A'$ and $\kappa_u'$ governing
the scaling of the distribution of avalanches for all forces below
$F_T$ are related by the same relations \kappaA and \kappau that hold
among the unprimed exponents,
\eqn\scalingp{\eqalign{
\kappa_A' &={\kappa '\over (1+\zeta)},\cr
\kappa_u' &=\kappa_A'+{1\over\zeta}(\kappa_A'-1).}}
The exponents obtained from our numerics satisfy well these scaling
relations.

The correlation length exponent $\nu\_$ can be
inferred from \primedexp\ or \kappaup\ if at least one of the
distribution at fixed driving force is computed numerically. This is
in general more difficult since the distribution is quite sensitive to
the value of the force increment used, which needs to be made very
small. We have evaluated numerically $D(S_\|;f)$ for three different
driving force. The distribution for $f=0.063$ is shown in Fig. 8.
We find $\kappa =0.05\pm 0.05$. Using the first of Eq.
\primedexp and the value $\kappa ' = 1.0\pm 0.2$ quoted earlier, we
obtain $\nu\_=1.13\pm 0.30$. The result is again consistent with
theoretical predictions. Finally, as a consistency check, we can now
use $\nu\_=1.13\pm 0.30$ in the second of Eqs. \primedexp
and Eq. \kappaup
to obtain $\kappa_A=0.07\pm 0.22$ and $\kappa_u=-0.95\pm 0.15$. These in
turn satisfy the scaling relationships \kappaA and \kappau\ .

The distribution of avalanches in discrete interface
growth models has been studied numerically by Sneppen \refs{\sneppen}
and by Sneppen and Jensen
\nref\jensen{K. Sneppen and M. H. Jensen, Phys. Rev. Lett. {\bf 71},
101(1993).}\refs{\jensen}. In their model the growing interface is
maintained in a ``critical state'' by a local growth rule similar to
that used in invasion percolation models that prevents the formation
of overhangs. The spatial and temporal correlations between successive
growth events and avalanches in this model have been studied
extensively by Leschhorn and Tang
\nref\leschhorn{H. Leschhorn and L-H Tang,
unpublished.}\refs{\leschhorn}. They find a rather complex behavior
where the distribution of growth events shows dynamical scaling only
locally. This is because in the Sneppen model the driving force is
self-tuned to maintain the interface in a ``critical state'' at the
onset of steady state motion, thereby introducing additional spatial
and temporal inhomogeneities in the model.

In contrast, in our continuous model the
distributions of various
measures of avalanche sizes displays a well defined dynamical scaling
characterized by a single correlation length $\xi\_\sim |f|^{-\nu\_}$
as the threshold is
approached from below. The correlation length exponent $\nu\_$ equals
the correlation length exponent $\nu$ obtained when the transition is
approached from above.
On the other hand, the values of the scaling
exponent of the avalanche size distribution, $\kappa\approx 0$, and of
the roughness exponent, $\zeta\approx 1$, indicate that the
interface may develop large local gradients, which were explicitly
excluded in our model. A physical interface will relax these large
strains by pinching off loops around the pinning centers, a mechanism
that may modify qualitatively the dynamics in the critical region both
below and above threshold. In particular this could reduced
considerably the value of the roughness exponent $\zeta$. Further
studies of a more general model that incorporates strong elastic
nonlinearities and allows for overhangs and loop generation are clearly
needed to address these questions.

\vskip .2in
This work was supported by the National Science Foundation through
Grants No. DMR91-12330 and DMR92- 17284. It was conducted using the
CM-2 at the National Center for Supercomputing Applications of the
University of Illinois at Urbana-Champaign and the CM-5 and SP-1 at
the Northeast Parallel Architecture Center at Syracuse University.
We thank Alan Middleton and Ofer Biham for many
useful discussions.

\vfill\eject
\listrefs

\vfill\eject

\figures

\fig{1}
{Creep velocity as a function of dimensionless temperature
$T$ at the $T=0$ threshold force $F_T$. The sample size is indicated.
The line shows the slope $\beta/\tau=0.16$ for the exponents values
discussed in the text.}

\fig{2}{Creep velocity as a function of dimensionless driving force
at various temperatures. The straight line shows the asymptotic
uniform velocity $v\sim F$. The data shown summarize results for three
different system size : $L=1024,4096,16384$. The same symbol is used
for different system size as no significant size dependence was
observed.}

\fig{3} {Scaled creep velocity $vT^{-\beta/\tau}$ versus scaled
reduced force $fT^{-1/\tau}$ for the three lowest temperatures of
Fig. 2, using $\tau=1.5$ and $\beta=0.24$.}

\fig{4} {Log-log plot of the energy barrier $U(F)$ defined as
$U(F)=T\log (F/v)$ versus driving force in the low temperature creep
region(from $T=3.16\times 10^{-4}$ to $T=1.00\times 10^{-2}$).
The dashed line has slope $-1/4$.}

\fig{5} {The area of avalanches obtained by successively stepping up
the driving force by small increments $f\approx 4.0\times 10^{-4}$ for
$-1\leq f\leq 0$ versus their diameter in the direction of the
interface for both $L=1024$ and $L=4096$. The straight line has slope 2.}

\fig{6} {The distribution of avalanche diameters (a) and areas (b)
integrated over all driving forces. The straight line has slope -2 for
(a) and $-{3\over 2}$ for (b).}

\fig{7} {The number distribution of displacement increments $\Delta
u$ for all driving forces below threshold for $L=4096$. The straight
line has slope -1.}

\fig{8} {The number distribution of avalanche diameters for fixed
driving force $f=-0.063$. The dimensionless $\Delta F$ used here is
$\Delta F\approx 7.0\times 10^{-4}$. The straight line has slope
$-1.05$.}

\end